\documentstyle[aas2pp4]{article}

\begin{document}

\title{FLOWS IN SUNSPOT PLUMES DETECTED WITH SOHO}

\author{N. Brynildsen, P. Maltby, P. Brekke, T. Fredvik, S. V. H. Haugan,\\
 O. Kjeldseth-Moe, and \O. Wikst\o l} 
\affil{Institute of Theoretical Astrophysics, University of Oslo, P.O. Box 1029 
Blindern, \\
0315 Oslo, Norway}

\begin{abstract}

Bright EUV sunspot plumes have been observed in eight out of 
eleven different sunspot regions with the Coronal Diagnostic Spectrometer 
-- CDS on SOHO.
>From wavelength shifts we derive the line-of-sight velocity, relative to 
the average velocity in the rastered area, 120$\arcsec \times$ 120$\arcsec$.
In sunspot plumes we find that the motion is directed away from the observer 
and increases with increasing line formation temperature, reaches a maximum 
between 15 and 41 km~s$^{-1}$ close to log T $\approx$ 5.5, then decreases 
abruptly.
The flow field in the corona is not well correlated with the flow in the 
transition region and we discuss briefly the implication of this finding. 

\end{abstract}


\section{Introduction}

Foukal et al. (1974) introduced the notation ``sunspot plumes'' to
describe areas above sunspot umbrae that are ``the brightest features 
in an active region by an order of magnitude''.  
This led to the idea that sunspot plumes are regions within large magnetic 
loops, extending to altitudes of several thousand kilometers above the 
photosphere, in which the temperature is one to two orders of magnitude 
lower than in the corona of the surrounding active region 
(Noyes et al. 1985).
In contrast, based on numerous sunspot observations with the Ultraviolet 
Spectrometer and Polarimeter (UVSP) on the {\it Solar Maximum Mission (SMM)} 
Gurman (1993) found that sunspot plumes were nearly nonexistent.
Most recently Maltby et al. (1998) observed sunspot plumes in five out 
of nine sunspot regions with the Coronal Diagnostic Spectrometer 
(CDS; Harrison et al. 1995) on the {\it Solar and Heliospheric 
Observatory (SOHO)} and discussed briefly previous conflicting results.
The CDS observations showed that sunspot plumes exist in 
the upper part of the transition region, occur both in magnetic 
unipolar and bipolar regions, and may extend outside the umbra and 
into the penumbra.

>From the energy requirements in sunspot loops Foukal (1976) suggested 
that rapid downflows occur in the plumes. 
Strong downflows over sunspots were reported by Brueckner, Bartoe, \&
VanHoosier (1977) and studied by Nicolas et al. (1982), while Kjeldseth-Moe 
et al. (1988) found that both upflows and downflows occurred.  Other
investigations have confirmed and extended these observations, for a
review see Maltby (1997). None of the observations above referred
specifically to plumes and to our knowledge the velocity in sunspot
plumes is not known. An investigation by Brynildsen et al. (1998) on
the connection between line emission and wavelength shift in sunspot
regions may, however, hold some relevance to this.
In this paper we extend the CDS material to twelve sunspot regions
and present the first measurements of velocities in sunspot plumes.

\section{Observations and Data Reduction}

Observations of twelve sunspot regions, listed in Table~1, were obtained 
with the Normal Incidence Spectrometer (NIS) of the CDS instrument 
as part of a joint observing program on {\it SOHO}. 
Since NOAA 7981 recurs as NOAA 7986 eleven different regions were observed.
A large fraction of the observing time was used to raster an area of 
120$\arcsec \times$ 120$\arcsec$, moving the narrow 2.0$\arcsec$ 
spectrometer slit perpendicular to the slit direction in steps of 2.0$\arcsec$.
The exposure time was 20~s; each raster required 25 min and contains 
information from 60 adjacent slit locations for ten emission lines. 
The observations contain spectra with spectral resolution,
$\lambda / \delta \lambda$, in the range 3635 -- 4500 within ten narrow 
spectral windows centered on the selected lines, which cover a wide range of 
ionization temperatures, see Table~2.
White light sunspot contours are taken from the MDI instrument 
(Scherrer et al. 1995) on SOHO and the National Solar Observatory.
To compensate for the influence of solar rotation on the alignment, 
the images were re-aligned by moving the frequently observed MDI magnetograms
artificially to the same observing time as the CDS recordings.

The data acquisition and detector characteristics that are relevant
for this study were described by Harrison et al. (1995).
Briefly, the CDS data are corrected for geometrical distortions, 
the CCD readout bias is removed, the non-wavelength-dependent 
calibration parameters peculiar to the detector are applied, including 
the exposure time, the amplification of the microchannel plate, and 
a flat-field correction.  
The final step in the calibration is to convert the photon events into 
absolute intensity units.

A careful approach to the line-of-sight velocity determination
is required to avoid the possible influence of other lines within 
the spectral windows, particularly in areas where the line of primary 
interest is weak.
Detailed investigations of the line profiles show that
most of them may be represented with a single Gaussian profile.
The profile parameters for He~{\scriptsize I} $\lambda$584 
and O~{\scriptsize III} $\lambda$599 are determined in this way 
since no disturbing lines are found.
We find that it is possible to improve the line-of-sight determination
for the other lines by representing the observations by a composite 
line profile, comprised of two Gaussian components, where one Gaussian 
is adjusted to fit the line of primary interest and the other  
accounts for the second most intense line within the spectral window,
see Table~2.
This approach is used even in cases where the distance between the
lines is too large to consider the second most intense line as a
blend.
After exploring different ways of reducing the data, we decided to use
two, slightly different methods.
In the first method the line profile parameters for both lines
are determined by a least squares fit to the observations.
This requires good signal to noise ratio both for the line and the blend
and is applied to the Mg~{\scriptsize IX} $\lambda$368 line.
In the second method we locate parts of the rastered area where 
the line of primary interest is weak and use this location to determine 
the wavelength position and line width of the second line.
Keeping these parameters constant, 
we are able to fit the observations with a composite line profile.
This method is tested by applying it to the He~{\scriptsize I} 
$\lambda$522 line and compare the results with those obtained 
for the strong He~{\scriptsize I} $\lambda$584 line, which is not 
influenced by other lines.
We estimate the accuracy in the line-of-sight velocity determinations
to be 5 km~s$^{-1}$ for He~{\scriptsize I} $\lambda$584 and 
O~{\scriptsize V}, 10 km~s$^{-1}$ for Mg~{\scriptsize IX}, and 
15 km~s$^{-1}$ for O~{\scriptsize III}, O~{\scriptsize IV}, 
Ne~{\scriptsize VI} and Fe~{\scriptsize XVI}.  

\section{Results}

Following Maltby et al. (1998) we set the criterion for the presence 
of a sunspot plume by requiring that the contours for peak line intensity 
$I \ge 5 \times \overline I$ are located (1) above the umbra or part 
thereof and (2) with most of the emission inside the white light sunspot.
Figure~1 (Plate L00) shows the observed spatial distribution of 
Ne~{\scriptsize VI} $\lambda$562 peak line intensity in twelve sunspots.
The brightest features with peak line intensity, $I$ larger than 5 times 
the average intensity, $\overline I$, are encircled by yellow contours,
whereas medium bright features with $I > 2.5 \times \overline I$ are 
encircled by green contours.
NOAA 7973, 7986, 8073, 8083, 8085, 8113 and 8123 satisfy the adopted
criterion for containing a sunspot plume.
We also regard NOAA 8011 as containing a sunspot plume since the 
O~{\scriptsize V} $\lambda$629 peak line intensity exceeds
5 $\times$ $\overline I$ in a small region above the umbra.
It should be remarked that sunspots without plumes, 
NOAA 7981, 7999, 8076 and 8108, also 
show Ne~{\scriptsize VI} $\lambda$562 peak line intensity 
$I > 5 \times \overline I$ both above and outside the sunspot.
Since NOAA 7981 recurs as 7986 we note that Figure~1 
(Plate L00) illustrates a finding of previous observers 
({\frenchspacing e.g.} Foukal 1976), that plumes may be absent 
during part of the sunspot's lifetime.

A remarkable feature in Figure~1 (Plate L00) is that the enhanced 
Ne~{\scriptsize VI} emission appears to outline one or a few thin emission
structures, extending from the sunspot to the surrounding regions.
In NOAA 8076 the extended feature resembles a magnetic loop, but in
other regions, such as NOAA 7986, the medium bright emission
features look like footpoints of magnetic loops. 
Similar features emitting strongly in the transition region lines
O~{\scriptsize IV} $\lambda$554 and O~{\scriptsize V} $\lambda$629
are observed in almost the same locations.
We note that Foukal (1976) remarked that strong cool emission was 
often found near both foot points of a loop, even if only 
one foot point could be confidently traced to the umbra.
This suggests that Foukal (1976) observed similar emission structures 
to those seen in Figure~1 (Plate L00).

We now consider the line-of-sight velocity in bright areas above sunspots,
{\frenchspacing i.e.} areas encircled with yellow contours in Figure~1 
(Plate L00) both in sunspot plumes and in equally bright areas above the 
other sunspots.
Figure~2 shows the relative line-of-sight velocity, $v$, versus line
formation temperature, $T$, in these bright areas above sunspots.
Since most of the sunspots are observed close to the disk 
centre we tend to use the words upflow and downflow, 
even though contributions from horizontal velocities cannot be excluded.
The CDS spectra contain few chromospheric lines and therefore 
the line-of-sight velocity is measured relative to the average
line-of-sight velocity in the rastered area, 
120$\arcsec \times$~120~$\arcsec$.
No corrections for the differential redshift between transition region- 
and chromospheric lines are applied.

Figure~2 shows that the relative line-of-sight velocity is 
directed away from the observer and increases with increasing temperature, 
reaches a maximum between 15 and 41 km~s$^{-1}$ close 
to log T $\approx$ 5.5 and then decreases abruptly.
For nine out of twelve sunspots the maximum relative velocity exceeds
26 km~s$^{-1}$.
The result is valid both for sunspot plumes and equally bright regions
above the other sunspots.
No connection between line-of-sight velocity and heliocentric angle, 
$\theta$, is apparent.
For the coronal lines with log T $\approx$ 6.0 (6.4) the velocity is 
below 10 km~s$^{-1}$ in eleven out of twelve sunspots. 
This implies a marked change from low velocities in the corona to
strong downflows in the sunspot transition region.
To clarify the problem let us study the spatial distribution of the
relative line-of-sight velocity in O~{\scriptsize V} $\lambda$629,
formed in the transition region and compare the results with those
obtained for the low corona line Mg~{\scriptsize IX} $\lambda$368,
see Figures~3 and 4 (Plates L00 and L00), respectively.
Figures~3 and 4 (Plates L00 and L00) confirm the results presented 
in Figure~2. 
Almost the entire area encircled with a yellow contour in the sunspot 
transition region is strongly redshifted, see Figure~3 (Plate L00),
whereas the corresponding areas in Figure~4 (Plate L00) show little or 
no wavelength shift. 
This fascinating result deserves a few comments.

The observations are obtained with small to moderate heliocentric angles
and show that the vertical flow in the corona is too small 
to maintain a strong flow in the sunspot transition region.
This suggests that the gas has to be supplied from regions surrounding the 
sunspot.
Let us examine Figures~3 and 4 (Plates L00 and L00).
Almost all sunspot regions in Figure~3 (Plate L00) contain one or a few 
prominent, strongly redshifted velocity channels, several of which extend 
from inside the sunspot to considerable distances from the sunspot.
We note that plasmas at transition region temperatures 
may occur at considerable heights within the active region 
({\frenchspacing e.g.} Brekke, Kjeldseth-Moe, \& Harrison 1997).
Since the gas is moving away from the observer in the velocity channels
that end in the sunspot, it seems likely that the gas is moving from 
regions located at a greater height outside the sunspots and towards the 
sunspot plume region.
This interpretation is compatible with the results presented in Figure~3 
(Plate L00) and with the low velocities observed in Figure~4 (Plate L00), 
where only NOAA 8076 shows a prominent, redshifted velocity channel. 
The present observations, interpreted in terms of gas at transition 
region temperature moving from greater height towards the sunspot 
plume, may be of interest in future studies of the suggestion that 
transition region structures are not physically connected to the 
coronal structures (Feldman 1983). 

It is interesting to compare corresponding images in Figures~1 and 3
(Plates L00 and L00).
We find that the enhanced line emission regions tend to be redshifted.
For a few sunspots, such as NOAA 8076, there is good correspondence between 
redshift and peak line intensity, whereas in others, such as NOAA 8113, 
there is a marked difference between the location of the enhanced line 
emission and the most prominent, redshifted channels.

Above we have tacitly assumed that the wavelength shift is caused by
material flow.
We note that the observations may be compatible with an alternative 
interpretation of redshifts in the transition region lines, 
proposed by Hansteen, Maltby, and Malagoli (1996).
Based on numerical simulations they find that episodic, magneto-hydrodynamic 
disturbances that originate in the corona and become nonlinear as they 
propagate towards the transition region may produce an overabundance of 
redshifts in the transition region lines, combined with small to moderate 
wavelength shifts in the coronal lines.

\acknowledgements 

We would like to thank all the members of the large international CDS 
team for their extreme dedication in developing and operating this 
excellent instrument, the Michelson Doppler Imager team 
and T. Rimmele at the National Solar Observatory for permission 
to use their data and the Research Council of Norway for financial support.
SOHO is a mission of international cooperation between ESA and NASA.

\clearpage

\begin{figure*}[p] 
 \figcaption[]{Observed spatial distribution of Ne~{\scriptsize VI} 
$\lambda$562 peak line intensity in twelve sunspot regions.
Enhanced intensities are shown as dark regions. 
Areas with peak line intensity, $I$, larger than 2.5 and 5 times the average 
intensity, $\overline I$, are encircled by green and yellow contours,
respectively. The images are oriented with north up, west to the right, and 
positions are given relative to the disk centre.
The umbral and penumbral contours are shown.}
\label{fig1}
\end{figure*}

\begin{figure*}[htbp]
\caption{Relative line-of-sight velocity, v,  versus line formation 
temperature, T, in the area above the sunspot where the Ne~{\scriptsize VI} 
$\lambda$562 peak line intensity $I > 5 \times \overline I$. 
The sunspots are ordered after increasing heliocentric angle, $\theta$, see
upper right hand corner for each sunspot.} 
\label{fig2}
\end{figure*}

\begin{figure*}[p] 
\caption{Spatial distribution of the relative line-of-sight velocity in 
O~{\scriptsize V} $\lambda$629. The velocities are measured relative to 
the average velocity in each image.
Motion towards (away from) the observer is shown with blue (red) color.
Green and yellow contours correspond to Ne~{\scriptsize VI} $\lambda$562 
peak line intensity, $I$~(Ne~{\scriptsize VI}), equal to 2.5 and 5 times 
the average intensity, $\overline I$~(Ne~{\scriptsize VI}).
For descriptions of positions, image orientation, and sunspot contours, 
see Figure~1.}
\label{fig3}
\end{figure*}

\begin{figure*}[p] 
\caption{Spatial distribution of the relative line-of-sight velocity in 
Mg~{\scriptsize IX} $\lambda$368. The velocities are measured relative to 
the average velocity in each image.
Motion towards (away from) the observer is shown with blue (red) color.
Green and yellow contours correspond to Ne~{\scriptsize VI} 
$\lambda$562~{\AA} peak line intensity, $I$~(Ne~{\scriptsize VI}), equal 
to 2.5 and 5 times the average intensity, $\overline I$~(Ne~{\scriptsize VI}).
White pixels mark regions without data or regions where the uncertainty
exceeds 10 km~s$^{-1}$.
For descriptions of positions, image orientation, and sunspot contours, 
see Figure~1.}
\label{fig4}
\end{figure*}

\clearpage

\begin{table}[htb]
\caption{{\scriptsize OBSERVED ACTIVE REGIONS}}
\begin{tabular}{llccc}
\tableline
     &                     &        &       &$\theta^{b}$ \\
NOAA & Date                & Days   & No.$^a$&(deg)    \\
\tableline
7973 & 1996 Jun 26         & 1 & 3 &  16     \\
7981 & 1996 Aug  2         & 1 & 5 &  16     \\   
7986 & 1996 Aug 29         & 1 & 3 &  17     \\    
7999 & 1996 Nov 27 - 29    & 3 & 9 & 19 - 46 \\   
8011 & 1997 Jan 16 - 17    & 2 &14 &  0 - 13 \\     
8073 & 1997 Aug 15-21      & 6 &33 & 10 - 41 \\        
8076 & 1997 Aug 29-31      & 3 &26 & 20 - 27 \\        
8083 & 1997 Sep  6 - 13    & 4 &31 & 35 - 64 \\        
8085 & 1997 Sep 15         & 1 & 4 &   44    \\        
8108 & 1997 Nov 18 - 21    & 2 &20 & 22 - 24 \\
8113 & 1997 Nov 29 - Dec 5 & 4 &45 & 19 - 43 \\
8123 & 1997 Dec 17 - 18    & 2 &16 & 20 - 32 \\
\end{tabular}
\tablenotetext{}{$^{a}$ No. = Number of rasters}
\tablenotetext{}{$^{b}$ The angle $\theta$ = heliocentric angle.}
\end{table}

\begin{table}[htb]
\caption{{\scriptsize OBSERVED EMISSION LINES}} 
\begin{center}
\begin{tabular}{lccllr}
\tableline
Selected & $\lambda$ & log T & Second  &$\lambda$ & \\
Line    &  ({\AA})  & (K)    &line$^{a}$&({\AA})   & \\
\tableline
He I    & 522.20 & 4.3      & Ne IV  & 521.8  & \\
He I    & 584.33 & 4.3      &        &        & \\
O III   & 599.60 & 5.0      &        &        & \\
O IV    & 554.51 & 5.2      & O IV   & 554.08 & \\
O V     & 629.73 & 5.4      & Ar VII & 630.30 & \\
Ne VI   & 562.80 & 5.6      & Ne VI  & 562.71 & \\
Mg VIII & 315.04 & 5.9      & Mg VI  & 314.6  & \\
Mg IX   & 368.07 & 6.0      & Mg VII & 367.7  & \\
Fe XIV  & 334.17 & 6.2      & N IV   & 335.05 & \\
Fe XVI  & 360.76 & 6.4      &        & 361.25 & \\
\tableline
\end{tabular}
\tablenotetext{}{$^{a}$ Second most intense line(s) within the spectral window}
\end{center}
\end{table}

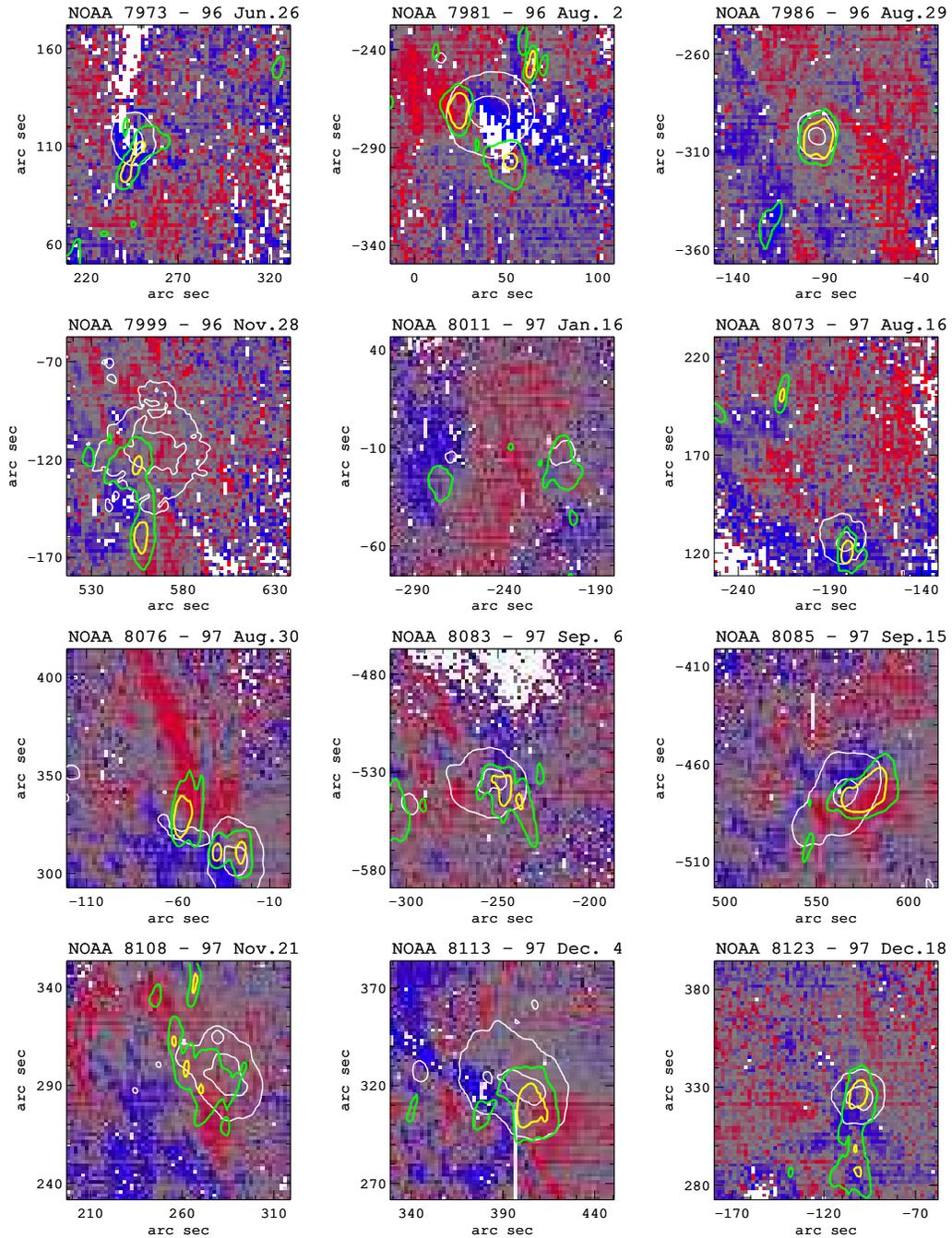
\end{document}